\documentstyle [12pt]{article}
\advance\voffset by -1 in
\advance\hoffset by -.5 in
\begin{document}
\def\p{\partial}
\def\a{\alpha}
\def\b{\beta}
\def\g{\gamma}
\def\tr{{\rm tr}\:}
\def\res{{\rm res}\:}
\def\tres{\tr\res}
\def\cA{{\cal A}}
\def\l{\lambda}
\def\L{\Lambda}
\def\f{\phi}
\def\hw{\hat{w}}
\def\hv{\hat{v}}
\def\t{t^*}
\def\Tau{{\rm T}}
\def\rarr{\rightarrow}
\def\bg{{\bf g}}
\def\d{\delta}
\def\s{\sum}
\def\op{^{\oplus}}
\def\om{^{\ominus}}
\def\tl{\tilde{\lambda}}
\def\tp{\tilde\p}
\def\tt{\tilde t}
\def\bof{{\bf f}}
\def\bb{{\bf b}}
\def\bh{{\bf h}}
\def\bc{{\bf c}}
\def\be{{\bf e}}
\def\F{\Phi}
\def\({\left(}
\def\){\right)}
\def\_k{_{k\a l}}
\centerline{{\bf Additional symmetries of the Zakharov-Shabat hierarchy,}} 
\centerline{{\bf String equation and Isomonodromy}}

\vspace{.5in}
\centerline{{\bf L.A.Dickey}}
\centerline{Dept. Math., University of Oklahoma, Norman, OK 73019
\footnote{e-mail: ldickey@math.ou.edu}}

\vspace{.3in}
We are going to explain and to prove the following statement:
{\em Isomonodromic deformations are nothing but symmetries of the 
Zakharov-Shabat (isospectral) hierarchy, both the basic ones (belonging to the
hierarchy) and additional, restricted to the submanifold of solutions to the 
string equation. All the symmetries commute with each other, so the 
equations can be integrated together.} 

A more precise formulation will follow (Theorem below). All necessary 
definitions will be given. We also construct quantities which are first 
integrals both for isospectral and for isomonodromic deformations.

Apparently, many formulas we are writing here can be extracted in one or 
another
form from the literature ([1-4]). What especially interested us was a context
and interpretation. Our goal is to express the theory in hierarchy
terms. It is common nowadays to study integrable systems not individually but
in big collections called hierarchies. A hierarchy is always a
set of commuting vector fields on an infinite dimensional
manifold.

A Zakharov-Shabat equation has a form $[\p_{t_1}-A_1(z),\p_{t_2}-
A_2(z)]=0$ where $A_1(z)$ and $A_1(z)$ are rational functions of $z$,
and $\p_{t_1}$ and $\p_{t_2}$ are derivatives with respect to some
variables. There are very many such equations. How to consider them as a
hierarchy, that was discussed in [5].  However, there is one more
activity where operators of type $\p_{t_i}-A_i(z)$ are involved. These are
the so-called monodromy preserving deformations, see [1-4].
Incorporating them into the ZS hierarchy, that is what we are doing 
here.\\

{\bf 1. Definition of the ZS hierarchy.} \\

A definition was presented  in [5], we repeat it here for convenience.
 
Let $a_k$, $k=1,...,N$ be a given set 
of complex numbers. Let, for every $k$, $$\hw_k=\sum_0^\infty w_{ki}(z-a_k)^i,
$$ be a formal series. The entries of $n\times n$ matrices $w_{ki}=\{
w_{ki,\a\b}\}$ belong to an algebra where det $w_{k0}$ is invertible. The 
formal series $\hw_k$ can be inverted. Define$$R_{k\a l}=\hw_kE_{\a}(z-a_k)^{
-l}\hw_k^{-1}$$ where $E_\a$ is a matrix with only one 
non-vanishing element, equal 1, on the $(\a,\a)$ place.

We have two kinds of objects. Such quantities as $\hw_k$ and $R_{k\a l}$ 
are formal series, or jets, at the points $a_k$. If $j_k$ is a
jet, then $j_k^-$ symbolizes its principal part, i.e., a sum of negative 
powers of $z-a_k$, and $j_k^+$ the rest of the series.
If the principal part contains finite number of terms it can be considered as 
a global meromorphic function. Global functions are objects of the second 
kind. A global function gives rise to a jet at every $a_k$, as its Laurent 
expansion there. In particular, $j_k^-$ can be
considered as a jet at a point $a_{k_1}$, different from $a_k$.\\

{\bf Definitions. (i)} {\sl A hierarchy corresponding to a fixed set $\{a_k\}$ 
is the totality of equations
$$\p_{k\a l}\hw_{k_1}=\left\{\begin{array}{l}-R_{k\a l}^+\hw_{k_1},~~k=k_1\\
~~R_{k\a l}^-\hw_{k_1},~~{\rm otherwise}\end{array}\right.,~~~\p_{k\a
l}=\p/\p t_{k\a l}.\eqno{(1)}$$ In the second case $R_{k\a l}^-$ is 
considered as a jet at $a_{k_1}$; $t_{k\a l}$ are some 
variables.}

{\bf (ii)} {\sl A ZS hierarchy is an inductive limit of hierarchies with fixed
sets $\{a_k\}$, with respect to a natural embedding of a hierarchy 
corresponding to a subset into a hierarchy corresponding to a larger set, as 
a subhierarchy.}
\\

Note that $R_{k\a0}=\hw_kE_{\a}\hw_k^{-1}=R_{k\a0}^+$ and $\p_{k\a0}\hw_{k_1}=
\hw_{k_1}E_\a;$ any linear combination of derivatives $\p_{k\a0}$
acting on $\hw$ multiplies it on the right by a constant diagonal matrix.
Nothing depends on this transformation, and we will not use the
variables $t_{k\a 0}$ at all.

Two jets $R_{k\a l}$ and $R_{k\b m}$ with the same $k$ commute.\\

{\bf Lemma.} {\sl Equalities $$\p_{k\a l}R_{k_1\a_1l_1}=\left\{\begin{array}{l}
-[R_{k\a l}^+,R_{k_1\a_1l_1}],~~k=k_1\\~~[R_{k\a l}^-,R_{k_1\a_1l_1}],~~{\rm 
otherwise}\end{array}\right.=[R_{k\a l}^-,R_{k_1\a_1l_1}]$$ hold.}\\

{\em Proof.} It easily can be obtained from the definition of $R_{k\a l}$.
$\Box$  \\

{\bf Proposition 1.} {\sl All operators $\p_{k\a l}$ commute.}\\

For a proof see [1]. This proposition entitles us to call the above defined
object a hierarchy.

The following proposition readily can be 
proven by a simple straightforward computation:\\

{\bf Proposition 2.} {\sl A dressing formula $$ \hw_{k_1}(\p_{k\a l}-E_{\a}
(z-a_k)^{-l}\d_{kk_1})\hw_{k_1}^{-1}=\p_{k\a l}-B_{k\a l},~~B_{k\a l}=
R_{k\a l}^-$$ is equivalent to Eq.(1).}\\

In the above equality, $B_{k\a l}$ is assumed to be a 
jet at $a_{k_1}$. However, it does not depend on $k_1$ at all and can be 
considered as a global function of $z$ with the only pole of $l$th order 
at $a_k$. Let $$w_k=\hw_k\exp\xi_k~~{\rm where}~\xi_k=\sum_{l=1}^\infty\sum_
{\a=1}^nt_{k\a l}E_{\a}(z-a_k)^{-l}.$$ 

{\bf Definition.} {\sl The collection $w=\{w_k\}$ is the formal Baker function
of the hierarchy.} \\

The statement of the Proposition 2 can be rewritten in terms of the Baker 
function as $$w_{k_1}\p_{k\a l}w_{k_1}^{-1}=\p_{k\a l}-B_{k\a l}\eqno{(2)}$$
and the equations (1) as $$\p_{k\a l}w_{k_1}=B_{k\a l}w_{k_1}.\eqno{(3)}$$

{\bf Proposition 3.} {\sl All operators $\p_{k\a l}-B_{k\a l}$ commute.}\\

This is an immediate corollary of the Proposition 1 and Eq. (2).\\

{\bf Remark.} Without difficulty we could include a case when a pole is 
at infinity. Then $\hw_\infty=\sum_0^\infty w_{\infty i}z^{-i}$. All formulas 
stay the same with $(z-a_k)$ being replaced by $z^{-1}$ and $+$ and $-$, as
superscripts, being swapped. For simplicity of writing, we skip this case.\\

One can consider arbitrary linear combinations of the above constructed 
operators,$$ L=\sum_{k\a l}\l_{k\a l}(\p_{k\a l}-B_{k\a l})=\p+U$$ 
where $\p=\sum_{k\a l}\l_{k\a l}\p_{k\a l}$ and $U=-\sum_{k\a l}\l_{k\a l}B_
{k\a l}$. Only finite number of $\l_{k\a l}$ are assumed to be non-zero. 
Two such operators commute which
yields equations of the Zakharov-Shabat type $$\p U_1-\p_1 U=[U_1,U].$$ 
Functions $U$ and $U_1$ are rational functions of the parameter $z$.\\  

{\bf 2. Additional symmetry and string equation.}\\

In the case of the KP hierarchy, it is well-known that there are
additional symmetries ([6],[7],[8], see also [9]) which do
not belong to the hierarchy, and do not commute between themselves. They
often are called ``Virasoro symmetries'', according to their rules of
commutation. Especially graphic and convenient is the way they are
presented in [7]. A similar construction can be performed for the
ZS hierarchy ([5]). Dressing an obvious relation $[\p_z,~\p\_k]=0$ with the 
help of $w_i$ at the point $a_i$ we have $0=w_i[\p_z,~\p_{k\a l}]w_i^{-1}=
[\p_z-M_i,\p_{k\a l}-B_{k\a l}],$ i.e.,
$$\p_{k\a l}M_i=\p_zB_{k\a l}-[M_i,~B_{k\a l}]\eqno{(5)}$$ where
 $$\p_z-M_i=w_i\p_zw_i^{-1}=\p_z-\p_zw_i\cdot w_i^{-1}=\p_z-\p_z\hw_i\cdot 
\hw_i^{-1}-\hw_i\xi_{iz}\hw_i^{-1}$$and $\xi_{iz}=\p\xi_i/
\p z= t_{i}\p_z\L_i$. The quantity $$M_i=
\p_z\hw_i\cdot\hw_i^{-1}+\hw_i\xi_{iz}\hw_i^{-1}\eqno{(6)}$$ is a jet at the 
point $a_i$. Notice that $$M_i^-=(\hw_i\xi_{iz}\hw_i^{-1})^-=(w_i\xi_{iz}w_i^
{-1})^- $$ while $B\_k$ can be written as $$B\_k=(w_k(\p\_k \xi_k)w_k^{-1})^
-.$$

Taking negative and positive parts of (5), we get at the point $a_i$ $$\begin
{array}{llll}(1)~i=k~~&\p_{k\a l}M_k^-
&=\p_zB_{k\a l}&-[M_k,~B_{k\a l}]_k^-\\ &\p_{k\a l}M_k^+&=&-[M_k,~B_{k\a l}]_k
^+\\(2)~i\neq k~~&\p_{k\a l}M_i^-&=&-[M_i,~B_{k\a l}]_i^-\\ &\p_{k\a l}M_i^+&=
\p_zB_{k\a l}&-[M_i,~B_{k\a l}]_i^+\end{array}\eqno{(7)}$$

{\bf Definition.} {\sl The additional symmetry is given by the system of 
differential equations}
$$\p^*w_j=(M_j^+-\sum_{i\neq j}M_i^-)w_j$$ where $\p^*$ is a
derivative with respect to a parameter $t^*$ of the symmetry. 

As it is easy to see, the equation of the additional symmetry 
implies $$\p^*R_{k\a l}=[M_k^+-\sum_{i\neq k}M_i^-,R_{k\a l}]~{\rm and}~
\p^*B_{k\a l}=[M_k^+-\sum_{i\neq k}M_i^-,B_{k\a l}].\eqno{(8)}$$
This has the meaning of the equality of two jets at $a_k$.\\

{\bf Proposition 4.} {\sl The additional symmetry commutes with operators of 
the hierarchy, $[\p^*,\p_{k\a l}]=0$, i.e., it is a symmetry, indeed.}\\

A proof is below. In the theory of the KP hierarchy
there is the celebrated string equation. It is nothing but condition of
invariance of field variables with respect to a definite additional symmetry 
(see, e.g., [10], [11]). In a more general sense, ``a string equation" 
is the invariance with respect to any additional symmetry.\\

{\bf Definition.} {\sl A string equation is condition that $w_i$ do not 
depend on $t^*$.} \\

This condition is compatible with the hierarchy, by virtue 
of the Proposition 4, and yields $$M_k^+-\sum_{i\neq k}M_i^-
=0,~~k=1,...,N. \eqno{(9)}$$

The jets $w_k$ are defined up to multiplication on the right by constant 
diagonal matrices that may be a series in powers of $(z-a_k)^{-1}$.
This does not change $R_{k\a l}$ and the equations of the hierarchy. However,
$M_k$ will modify. One can use that possibility to incorporate terms of 
degree $-1$ into $\p_z\xi_k$, so put $\xi_k=\sum_{l=1}^\infty\sum_
{\a=1}^nt_{k\a l}E_{\a}(z-a_k)^{-l}+\sum_{\a=1}^n\l_{k\a}\log(z-a_k)$
to have $$\p_z\xi_k=-\sum_{l=1}^\infty\sum_
{\a=1}^nt_{k\a l}E_{\a}l(z-a_k)^{-l-1}+\sum_{\a=1}^n\l_{k\a}(z-a_k)^{-1}.$$

{\bf 3. Toward the isomonodromy equations.}\\

Let us introduce new (``additional") variables $\t_i$ and write
differential equations $$\p_i^*w_j=\left\{\begin{array}{rr}-M_i^-w_j,&
i\neq j\\M_i^+w_j,&i=j\end{array}\right.,~\p_i^*=\p/\p\t_i,$$the left-
and the right-hand sides are jets at $a_l$. A sum of
these vector fields equals the vector field $\p^*$, the additional
symmetry, $$\p^*w_j=\sum_i\p_i^*w_j.$$
 Is each of them a symmetry? Alas, this is not the case.
Nevertheless, they also have an advantage: they commute between
themselves.\\

{\bf Proposition 5.} {\sl The following commutation rules hold:}
$$[\p_i^*,~\p\_k]w_j=(\d_{ik}-\d_{ij})(\p_zB\_k)w_j,~~[\p_i^*,~\p_k^*]w_j=0.$$

{\em Proof.} In order to prove the first relation, we consider the
cases when all three indices $i,k$ and $j$ are distinct, when a pair of
them coincide, and when all three are equal. Different letters will
always symbolize different numbers here.

1)$$ \p_i^*\p\_k w_j=\p^*_i(B\_k w_j)=-[M_i^-,B\_k]^-_k w_j-B\_k M_i^-w_j.$$
Let us explain. $M_i^-$ is a rational function with a pole at $a_i$,
we consider it as a global function generating a (positive) Laurent series at
$a_k$. Then we take its commutator with $B\_k$ which is a negative jet
at this point and take the negative part of the commutator. It
is a global function having a purely positive expansion at $a_j$.
A superscript $-$ accompanied by a subscript $k$ denotes taking
the negative part of an expansion in powers of $z-a_k$.

How we got the formula? Compute: $$\p_i^*B\_k=\p_i^*(w_k(\p\_k \xi_k)w_k^{-1})_k
^-$$ $$=-(M_i^-w_k(\p\_k \xi_k)w_k^{-1}-w_k(\p\_k \xi_k)w_k^{-1}M_i^-)_i^-
=-[M_i^-,R\_k]^-_k=-[M_i^-,B\_k]^-_k.\eqno{(10)}$$ 
The matrix $M_i^-$ has a purely positive expansion at $a_k$, therefore
the positive part of $R\_k$ does not make any contribution to the
negative part of the commutator, and one can replace $R\_k$ by $B\_k$. Now,
$$\p\_k\p_i^*w_j=-\p\_k(M_i^-w_j)=-[B\_k,M_i^-]^-_iw_j-M_i^-B\_k w_j.$$ The
formula $$\p\_k M_i^-=[B\_k,M_i^-]^-_i \eqno{(11)}$$ used here immediately 
follows from (7). Finally, $$[\p_i^*,\p\_k]w_j=-([M_i^-,B\_k]^-_k
+[M_i^-,B\_k]^-_i)w_j-[B\_k,M_i^-]w_j=0$$ since a sum of all the principal
parts of a rational function equals the function itself (generally, up
to a constant, the latter is zero here because the function decays at
infinity).

2) It is easy to see that in the previous case it was not important that
$k\neq j$, i.e., $$[\p_i^*,\p_{j\a l}]w_j=0.$$

3) $$\p_k^*\p\_k w_j=\p_k^*(B\_k w_j)=[M_k^+,B\_k]^-_kw_j-B\_k M_k^-w_j$$ and
(see the first of eqs. (7))
$$\p\_k\p_k^*w_j=-\p\_k(M_k^-w_j)=-(\p_zB\_k-[M_k,B\_k]_k^-)w_j-M_k^-B\_k w_l$$
whence $$[\p_k^*,\p\_k]w_j=-([M_k^-,B\_k]^-_k+[B\_k,M_k^-]-\p_zB\_k)w_j=
(\p_zB\_k)w_j.$$

4) $$\p_j^*\p\_k w_j=\p_j^*(B\_k w_j)=-[M_j^-,B\_k]^-_kw_j+B\_k M_j^+w_j$$ $$
=-(-[M_j^-,B\_k]+[M_j^-,B\_k]_j^-)w_j-B\_k M_j^+w_j   
=([M_j^-,B\_k]-[M_j,B\_k]_j^-)w_j-B\_k M_j^+w_j$$ and
(see the last of the eqs. (7))
$$\p\_k\p_j^*w_j=\p\_k(M_j^+w_j)=-(\p_zB\_k-[M_j,B\_k]_j^+)w_j-M_j^+B\_k w_j.$$
Therefore, $$[\p_j^*,\p\_k]w_j=-([M_j^-,B\_k]-[M_j,B\_k]+[M_j^+,B\_k]+
\p_zB\_k)w_j=-(\p_zB\_k)w_j.$$

5) $$\p_{j\a l}^*\p_{j\a l}w_j=\p_j^*(B_{j\a l}w_j)=[M_j^+,B_{j\a l}]^-_jw_j+
B_{j\a l}M_j^+w_j$$ and (see the second of the eqs. (7))
$$\p_{j\a l}\p_j^*w_j=\p_{j\a l}(M_j^+w_j)=-[M_j,B_{j\a l}]^+w_j+M_j^+B_{j\a l}
w_j=-[M_j^+,B_{j\a l}]^+w_j+M_j^+B_{j\a l}w_j$$ whence
$$[\p_j^*,\p_{j\a l}]w_j=[M_j^+,B_{j\a l}]w_j-[M_j^+,B_{j\a l}]w_j=0.$$ 

The first assertion is proven. The second one requires the following
computations.

6) $$\p_k^*\p_i^*w_j=-\p_k^*(M_i^-w_j)=[M_k^-,M_i]_i^-w_j+M_i^-M_k^-w_j
=[M_k^-,M_i^-]_i^-w_j+M_i^-M_k^-w_j$$ and
$$\p_i^*\p_k^*w_j=[M_i^-,M_k^-]_k^-w_j+M_k^-M_i^-w_j$$
which implies
$$[\p_k^*,\p_i^*]w_j=([M_k^-,M_i^-]_i^-+[M_k^-,M_i^-]_k^--[M_k^-,M_i^-
])w_j
=0.$$

7) It is easy to see that nothing changes in the preceding poof if
$k=j$. Thus, $$ [\p_k^*,\p_i^*]w_j=0$$ for all $k,i$ and $j$.
The proposition is proven.\\

{\bf Corollary.} {\sl The sum $\p^*=\s \p_i^*$ commutes with all $\p\_k$.
This is the above additional symmetry.}\\

Indeed, $$[\s\p_i^*,\p\_k]w_j=(\p_zB\_k-\p_zB\_k)w_j=0.$$

The Baker function $\{w_j\}$ depends on $\{a_i\}$ as parameters.
The next important step is to make the parameters variable, namely,
put $a_i=t^*_i$. The total derivative with respect to $a_i$ is
$D_i=\p_i^*+\p/\p a_i$ where $\p/\p a_i=\p_{a_i}$ is a partial derivative with 
respect to $a_i$ which enters $w_j$
explicitly in the form of $z-a_i$. Evidently, $(\p/\p a_j)w_j=-\p_zw_j$ and
$(\p/\p a_i)w_j=0$ when $i\neq j$, hence
$$D_iw_j=-M_i^-w_j,~i\neq j;~D_jw_j=(M^+_j-M_j)w_l=-M_j^-w_j,$$ i.e., in
all cases $$D_iw_j=-M_i^-w_j.\eqno{(12)}$$

{\bf Proposition 6.}{\sl Vector fields (12) commute with each other and with
$\p\_k$'s.}\\

{\em Proof.} We have,
$$D_kD_iw_j=-D_k(M_i^-w_j)=[M_k^-,M_i]_i^-w_j+M_i^-M_k^-w_j=
[M_k^-,M_i^-]_i^-w_j+M_i^-M_k^-w_j$$ and
$$D_iD_kw_j=-\p_i^*(M_k^-w_j)=[M_i^-,M_k^-]_k^-w_j+M_k^-M_i^-w_j,$$
hence $$[D_k,D_i]w_j=([M_k^-,M_i^-]_i^-+[M_k^-,M_i^-]_k^--[M_k^-,M_i^-]_i)
w_j=0.$$ Since the commutators $[\p_i^*,\p\_k]$ are already found
(proposition 5), it remains to find additional terms $[\p_{a_i},\p\_k]w_j$.
It is easy to see that this is zero when all three indices are distinct.
Now, $$\p_{a_k}\p\_k w_j=\p_{a_k}(B\_k w_l)=-(\p_zB\_k)w_j,~\p\_k\p_{a_k}w_j
=0,$$ Thus, $[\p_{a_k},\p\_k]w_j=-(\p_zB\_k)w_j$. This additional term exactly
cancels with the value of the commutator in the proposition 5.

Similarly, $$\p_{a_j}\p\_k w_j=\p_{a_j}(B\_k w_j)=B\_k(-\p_zw_j)=-
B\_k M_jw_j,$$ and $$\p\_k\p_{a_j}w_j=-\p\_k(M_jw_j)=-(\p_zB\_k-
[M_j,B\_k])w_j-M_jB\_k w_j$$ (see (5)). This yields that $$[\p_{a_j},\p\_k]w_j=
(\p_zB\_k)w_j$$ which also cancels. Finally, $$\p_{a_j}\p_jw_j=
\p_{a_j}(B_{j\a l}w_j)=-\p_z(B_{j\a l}w_j)$$and $$\p_{j\a l}\p_{a_j}w_j=
\p_{j\a l}(-\p_zw_j)=-\p_z(\p_{j\a l}w_j)=-\p_z(B_{j\a l}w_j)$$ and 
$[\p_{a_j},\p_j]w_j=0$. This completes the proof.\\

{\bf Corollary.} {\sl If (12) is considered as a system of differential
equations where $D_i$ are total derivatives with respect to $a_i$ then
this system is consistent and also compatible with the hierarchy equations
$\p\_k w_j=B\_k w_j$.}\\ 

Thus, $D_i$ are true symmetries of the ZS hierarchy, in contrast to $\p_i^*$. 
We call them {\em additional symmetries}\footnote{Commutativity of additional
symmetries seemingly contradicts the statement that they are analogues of the
nonkommutative Orlov-Shulman additional symmetries. In fact the whole collection
of $D_i$ is an analogue of only one O-S symmetry which splits into N commuting 
symmetries, accordingly to the number of poles.}.  

Now, we turn to the additional symmetry $\p^*=\sum\p_i^*$. In terms of
$D_i$ that is $$\p^*=\p_z+\sum D_i.$$ This operator commutes with all $D_k$'s 
since $\p_z$ commutes with all other involved vector fields. The string 
equation $(\p_z+\sum D_i)w_j=0$ is, therefore, invariant with respect to the 
equations (12). In more detail, the string equation looks like
$$(\p_z-\s M_i^-) w_j=0,~~j=1,...,N. \eqno{(13)}$$

Summarizing everything said before, we formulate the following\\

{\bf Theorem.} {\sl Let $$\xi_k=\sum_{l=1}^\infty\sum_
{\a=1}^nt_{k\a l}E_{\a}(z-a_k)^{-l}+\sum_{\a=1}^n\l_{k\a}\log(z-a_k)$$
where $\{t\_k\},\{a_k\}$ and $z$ are independent complex variables. 
Let $\p\_k=\p/\p t\_k$ and $D_k=\p/\p a_k$ be ``total partial
derivatives" taking into account variables involved both explicitly and
implicitly through the unknowns $w_{k}$. 
Let $$w_k=\s_0^\infty w_{km}(z-a_k)^m\exp\xi_k,~B\_k=(w_k(\p\_k \xi_k)w_k^{-
1})^-,~M_k^-=(w_k(\p_z\xi_k)w_k^{-1})^-$$ where the superscript $-$ means the 
negative (principal) part of the formal Laurent series in powers of $z-a_k$.
Then all equations:
$$(\p\_k-B\_k)w_j=0,~k,j=1,...,N;~\a=1,...,n,~l=1,...\eqno{(14)}$$
$$(D_k+M_k^-)w_j=0,~k,j=1,...,N,\eqno{(15)}$$ and
$$(\p_z-\s_k M_k^-)w_j=0,~j=1,...,N \eqno{(16)}$$ are compatible. Thus,
they can be solved all together and a solution $w_j(t;a;z)$ 
found.}\\ 

The whole system is called {\em isomonodromic}\footnote{{\rm About
the isomonodromic deformations see [1], [2], [3],
[4].}} equations, Eqs. (14) describing monodromy preserving deformations of 
diagonal elements $t\_k$ and Eqs. (15) those of poles. The equations (14) 
(without restriction (16)) are known as the ZS {\em isospectral} hierarchy. \\

{\bf 4. First integrals.}\\

Let $$R_i^C=w_iCw_i^{-1}$$ where $C$ is a constant diagonal matrix. \\

{\bf Proposition 7.} {\sl The jets at the point $a_i$:
$$J_{ik\a l}(z)=\int\tr R_i^C\p_zB\_k dt_{k\a l}~{\rm and}~J_{ik}^*=\int\tr
R_i^C\p_zM_k^-da_k,~~\forall k$$ are generators of first integrals of
the equtions (14) and (15), i.e.,
coefficients of their expansions in powers of $z-a_i$ are first integrals.}\\

The integral is understood in an algebraic sense: if $f(t)$ is a differential 
polynomial in elements $\{w_{jm,\a\b}\}$ then $\int f(t)dt\_k$ is the 
class of equivalence of $f(t)$ modulo exact derivaties $\p\_k g(t)$ where $g(t)
$ are also differential polynomials.

{\em Proof.} (1) For brevity, we write $B_k$ instead of $B\_k$, $\p_k$ for 
$p\_k$, and so on.  

We have $$\p_l\:\tr R_i^C\p_zB_k=\tr([B_l,R_i^C]\p_zB_k+
R_i^C\p_z[B_l,B_k]_k^-).$$ This is an equality of jets at $a_i$. Now we
use the following transformations: $$\tr[B_l,R_i^C]\p_zB_k=-\tr
R_i^C[B_l,B_{k,z}]~{\rm where}~B_{k,z}=\p_zB_k,~~[B_l,B_k]=[B_l,B_k]_k^-
+[B_l,B_k]_l^-$$ and $[B_l,B_k]_l^-=-\p_kB_l$ whence
$$\p_l\:\tr R_i^C\p_zB_k=\tr(-R_i^C[B_l,B_{k,z}]+R_i^C\p_z[B_l,B_k]+
R_i^C\p_kB_{l,z})=\tr(R_i^C[B_{l,z},B_k]+R_i^C\p_kB_{l,z})$$ $$=
\tr([B_k,R_i^C]B_{l,z}+R_i^C\p_kB_{l,z})=\tr((\p_kR_i^C)B_{l,z}+R_i^C\p_kB_{l,z
})=\p_k\:\tr R_i^CB_{l,z}.$$ A nice equality is obtained: $$\p_l\:\tr R_i^C\p_z
B_k=\p_k\:\tr R_i^C\p_zB_l.$$ It implies that $\p_l\int\tr R_i^C\p_zB_kdt_k=0$.

(2) Now, we compute $D_l\:\tr R_i^C\p_zB_k$. First, let $l\neq k$.
$$D_l\:\tr R_i^C\p_zB_k$$ $$=\tr(-[M_l^-,R_i^C]B_{k,z}-R_i^C[M_l^-,B_k]_{k,z}^-
)=\tr(-[M_l^-,R_i^C]B_{k,z}-R_i^C[M_l^-,B_k]_z+R_i^C[M_l^-,B_k]_{l,z}^-)$$
Transform the last term using (7): $$R_i^C[M_l^-,B_k]_{l,z}^-=
R_i^C[M_l,B_k]_{l,z}^-=R_i^C(-\p_kM_{l,z}^-).$$ We have $$D_l\:\tr R_i^C\p_zB_k
$$ $$=\tr(-R_i^C[M_{l,z}^-,B_k]-R_i^C\p_kM_{l,z}^-)=\tr(-M_{l,z}^-\p_kR_i^C-
R_i^C\p_kM_{l,z}^-)=-\p_k\:\tr (R_i^CM_{l,z}^-).$$ 

Secondly, let $l=k$. We have $$D_lB_l=(\p_l^*+\p_{a_l})B_l=[M_l^+,B_l]-
\p_zB_l=[M_l^+,B_l]-(\p_lM_l^-+[M_l,B_l]_l^-)$$ (see (7)), therefore,
$$D_l\:\tr R_i^C\p_zB_l$$ $$=\tr(-[M_l^-,R_i^C]B_{l,z}
+R_i^C[M_l^+,B_l]_{l,z}^--R_i^C(\p_lM_{l,z}^-+[M_l,B_l]_{l,z}^-)
)$$ $$=\tr(R_i^C[M_l^-,B_{l,z}]-R_i^C[M_l^-,B_l]_{l,z}^--
R_i^C\p_lM_{l,z}^-)$$ $$=\tr(-R_i^C[M_{l,z}^-,B_l]_l^--
R_i^C\p_lM_{l,z}^-)=\tr(-M_{l,z}^-(\p_lR_i^C)-R_i^C\p_lM_{l,z}^-)$$
$$=-\p_l\:\tr(R_i^CM_{l,z}^-).$$ Thus, we obtained two formulas
$$\p_l\:\tr R_i^C\p_zB_k=\p_k\:\tr R_i^C\p_zB_{l},~{\rm and}~~D_l\:\tr R_i^C
\p_zB_k=-\p_k\:\tr R_i^C\p_zM_{l}^-\eqno{(17)}.$$

One easily can prove one more equation, in addition to Eqs. (17),
$$ D_k\:\tr R_i^CM_{l,z}^-=D_l\:\tr R_i^CM_{k,z}^-\eqno{(18)}$$  which 
completes the proof.

Together, (17) and (18) are equivalent to the statement that the form
$$\omega=\tr\sum\_k R_i^CB_{k\a l,z}dt\_k-\tr\sum_k R_i^CM_{k,z}^-da_k$$ is 
closed.\\

{\bf Remark.} The relations (17) and (18) hold always, whenever $l$ equals
$i$ or not. However, if one wishes to obtain conservation laws expanding
(17) and (18) in powers of $z-a_i$, the case $l=i$ will be exceptional,
one fails to get a conservation law. It is better 
to eliminate this case demanding that $a_i$ be an additional point that is 
not variable. For example, it is convenient to take this point at infinity.\\

{\bf References.}\\

1. Flaschka, H. and Newell, A.: Monodromy- and spectrum-preserving
deformations I. Comm. Math. Phys. 76, 65-116 (1980)\\

2. Jimbo, M., Miwa, T. and Ueno, K.: Monodromy preserving deformation of
linear ordinary equations with rational coefficients. I, Physica 2D, 
306-352 (1981). II, Physica 2D, 407-448 (181). III  Physica 4D, 26-46
(1981)\\

3. Ueno K.: Monodromy preserving deformation of linear differential
equations with irregular singular points. Proc.  Japan Acad. Sc. 56,
ser.A, 97-102 (1980)\\

4. Harnad, J. and Its, A.: Integrable Fredholm operators and dual
isomonodromic deformations. Preprint solv-int/9706002 (1997)\\

5. Dickey, L.A.: Why the general Zakharov-Shabat equations form a
hierarchy. Comm. Math. Phys. 163, 509-521 (1994)\\

6. Chen, H.H., Lee, Y.C. and Lin, J.F.: On a new hierarchy of symmetries
for the Kadomtsev-Petviashvili equation. Physica 9D, 439-445 (1983)\\

7. Orlov, A.Yu. and Shulman, E.I.: Additional symmetries for integrable
equation and conformal algebra representations. Lett. Math. Phys. 12,
171-179 (1986)\\

8. Fokas, A.S.: Symmetries and integrability. Studies in Appl. Math. 77,
253-299 (1987)\\

9. Dickey, L.A.: (a) {\em Soliton Equations and Hamiltonian Systems.} Adv. 
Series in  Math. Phys. 12, World Scientific (1991)\\
(b) Lectures on classical W-algebras. Acta Applicandae Mathematicae 47, 
243-321 (1997)\\

10. (a) Adler, M. and van Moerbeke, P.: A matrix integral solution to 
two-dimensional gravity. Commun. Mat. Phys. 147, 25-56 (1992)\\
(b) van Moerbeke, P.: Integrable foundations of string theory. In:
{\em Lectures on Integrable Systems}, eds. Babelon, O. et all. World 
Scientific (1994)\\

11. Dickey, L.A.: Additional symmetries of KP, Grassmannian, and the
string equation. Modern Phys. Lett. A 8, 1259-1272 (1993)\\

\end{document}